\documentclass[useAMS,usenatbib]{mn2e}
\usepackage{epsfig}
\usepackage{amsmath}
\usepackage{amssymb}

\newcommand{\sinc}{{\rm sinc}}

\title{ Angular correlation of the stellar scintillation on large telescopes}
\author[V.~Kornilov]{V.~Kornilov\thanks{E-mail: victor@sai.msu.ru}\\
Lomonosov Moscow State University, Sternberg Astronomical Institute, Universitetsky prosp. 13, 119234 Moscow, Russia}

\begin{document}

\date{Accepted 2012 June 12.  Received 2012 June 11; in original form 2012 March 29}

\pagerange{\pageref{firstpage}--\pageref{lastpage}}

\pubyear{2012}

\maketitle
\label{firstpage}

\begin{abstract}
The stellar scintillation is one of the fundamental limitation to the precision of ground-based photometry. The paper examines the problem of correlation of the scintillation of two close stars at the focus of a large telescope. The derived correlation functions were applied to data of the long-term study of the optical turbulence (OT) in the Northern Caucasus with MASS (Multi-Aperture Scintillation Sensor) instrument to predict the angular correlation of the scintillation at the Sternberg institute 2.5~m telescope currently in construction. A median angular radius of  the correlation as large as 20 arcsec was found for the case of Kolmogorov OT.
On the basis of the obtained relations we also analyze the correlation impact in ensemble photometry and conjugate plane photometry. It is shown that a reduction of the scintillation noise up to 8 times can be achieved when using a crowded ensemble of comparison stars. The calculation of the angular correlation can be repeated for any large telescope at the site where the OT vertical profiles are known.
\end{abstract}


\section{Introduction}

Many astronomical problems require high-precision photometric data. Two clear examples are the determination of parameters of extraterrestrial planetary systems from eclipsing light curves \citep{Everett2001,Winn2009,Southworth2009}, and the study of low-amplitude stellar pulsations, reflecting inner structure of stars \citep{Heasley1996,Aerts2010book}. The success of the CoRoT and Kepler space missions clearly demonstrates the need for further increase of the accuracy of astronomical photometry.

Despite the superiority of space projects, observations with ground-based telescopes also have some advantages. They are more accessible, the number of the telescopes is large and their service life is tens times larger than for space missions. The larger apertures of these telescopes push the intrinsic limit of the accuracy, caused by photon noise,  towards fainter objects or improved temporal resolution.

The main limitation to obtain high-precision photometric results from the ground is due to the Earth's atmosphere, in particular the atmospheric scintillation of stars. The stellar scintillation is a random temporal fluctuations of the radiation entering the telescope aperture and is related to the amplitude distortions of the light wave having passed through the turbulent atmosphere. Photometric errors induced  by the scintillation ({\it scintillation noise}) have repeatedly been studied \citep[see, e.g.,][]{Young1969,Dravins1997a}.

In comparison to the photon noise, the scintillation noise does not depend on the brightness of the stars and thus cannot be reduced by choosing brighter objects. For typical exposure times (few seconds and longer), the scintillation noise decreases more slowly with telescope diameter than the photon noise and thus becomes to dominate on large telescopes. Consequently exactly the scintillation noise sets a limit to the photometric accuracy \citep{Heasley1996}. This problem becomes even more relevant for exposures shorter than 1~s.

In contrast to the flux variations due to changes in atmospheric transparency, the stellar scintillation is characterized by a lower spatial (angular) correlation \citep{Dravins1997a}. That leads to an increase of the errors in differential photometry when using a distant comparison star. It is assumed that the angular correlation for the scintillation is typically a few arcseconds  but the special case of large telescopes remains virtually unexplored.

This paper is a theoretical study of the angular correlation of the stellar scintillation on the basis of researches on the scintillation as a tool for probing OT above astronomical sites. We first consider the link between OT parameters and photometric scintillation noise. Then the formulas for the correlation at large telescopes in two limiting regimes, short (much less than 1~s) and long (greater than 1~s) exposures, are derived and analyzed.

In Section~\ref{sec:prognoz}, the results are applied to the OT vertical distribution above Mt Shatdzhatmaz in the North Caucasus \citep{2010MNRAS} where the Sternberg Astronomical Institute (SAI) 2.5~m diameter telescope is to be installed. The angular correlation functions and typical angles of the correlation are calculated. Finally, the results are applied to two methods aimed to reduce the scintillation noise, the recently proposed photometry in a conjugated entrance pupil \citep{Osborn2010} and the more traditional ensemble photometry \citep{Gilliland1993}.

\section{Theory}
\label{sec:theory}

\subsection{Scintillation noise in photometry}

In differential photometry, the light flux from the science object $1$ is compared to the flux from another reference object $2$ in order to remove the uncertainties caused by instrumental factors and by variations in atmospheric transparency. The comparison can be performed as the ratio of $I_1/I_2$ but more often the difference between stellar magnitudes: $\Delta m = 1.086\,\ln I_1/I_2$.

Several noise sources of different nature affect these measurements. Depending on its origin, the noise enters in the measured signal additively (detector noise, sky background noise), multiplicatively (atmospheric transparency fluctuations and scintillation noise) or in a more complex way (photon noise). Hereafter, we consider only one component, the scintillation noise $\delta I$ which is proportional to the mean light intensity or its equivalent in magnitudes,  the additive noise $\delta m$. By definition, the mathematical expectation $\langle \delta I \rangle \equiv 0$ and $\langle \delta m \rangle \equiv 0$.

Generally accepted that the power of the stellar scintillation is measured with the {\it scintillation index}
which is defined as $s^2 = \langle \delta I^2 \rangle/\langle I \rangle^2$ or $s^2 = \langle (\ln I - \ln \langle I \rangle)^2 \rangle$. For small fluctuations of flux, these definitions are equivalent.

Therefore, we can present the noise variance (scintillation noise power) in the form as $\sigma^2_I = \langle I \rangle^2 s^2$ for intensities, or $\sigma^2_m = 1.179\,s^2$ for magnitudes. The normalised variance of the ratio $o = I_1/I_2$ and the variance of the magnitudes difference $\sigma^2_{\Delta m}$ are simply expressed in the terms of the scintillation indices:
\begin{equation}
\frac{\sigma^2_{o}}{o^2} = \frac{\sigma^2_{\Delta m}}{1.179} = s^2_1 + s^2_2 - 2\, c_{1,2},
\label{eq:mag_index}
\end{equation}
where $c_{1,2}$ is the {\it cross-index}, defined naturally as
\begin{equation}
c_{1,2} = \frac{\langle \delta I_1 \, \delta I_2 \rangle}{\langle I_1 \rangle \langle I_2 \rangle}.
\end{equation}
When two objects are measured simultaneously in the same field of view, we can assume that their scintillation is equal: $s^2_1 = s^2_2$, and the cross-index depends only on the angular distance $\theta$ between the objects. Thus, the previous formula (\ref{eq:mag_index}) can be written as $s^2_d = 2\,s^2 - 2\,c(\theta)$.

For uncorrelated signals, $s^2_d$ is twice as large as the scintillation index, i.e. the scintillation noise will be doubled after the comparison. On the contrary, in the case of a significant correlation, the scintillation noise is reduced. This is a very important fact which, therefore, has been examined in both theoretical \citep{Ryan1998} and practical \citep{Dravins1997a} aspects.

It seems reasonable to define the {\it angular correlation radius} $\theta_1$ as the angle at which $s^2_d$ becomes equal to $s^2$, i.e. when there is no amplification of the scintillation noise after to comparison. Obviously then $c(\theta_1) = s^2/2 = c(0)/2$, so that the parameter $\theta_1$ is exactly a half-width of the main correlation peak.

\subsection{Basic relations}

It is known \citep[see, e.g.,][]{Roddier1981}, that under the approximation of weak perturbations, the scintillation index measured at the Earth's surface, is defined by the distribution of the structural coefficient $C_n^2(z)$ along the line of sight throughout the whole atmosphere using the following equation:
\begin{equation}
s^2 = \int_A C_n^2(z)\,Q(z) \,{\rm d}z,
\label{eq:s2int}
\end{equation}
where $z$ is the distance of the turbulent layer. For a measurement in the zenith, the distance equals to the layer altitude $h$ above the site. Similarly one can define the cross-index (covariance):
\begin{equation}
c(\theta) = \int_A C_n^2(z)\,R(z,\theta) \,{\rm d}z,
\label{eq:cint}
\end{equation}
The $Q(z)$ and $R(z,\theta)$ are the {\it weighting functions} relating the OT intensity inside the layer to its effect in the observing plane. Small-amplitudes approximation works always for large apertures, so this assumption is not restrictive for what follows.

The technique for computing the $Q(z)$ was described previously \citep[see, e.g.,][]{Tokovinin2002b,Tokovinin2003} in details. For Kolmogorov model of isotropic and locally homogeneous turbulence with 2D spatial power spectrum of wave phase perturbations $\propto f^{-11/3}$ (where $f$ is the modulus of the spatial frequency), there is a known expression for measurement with zero exposure time:
\begin{equation}
Q(z) = 9.61\int_0^\infty f^{-11/3}\,S(z,f) A(f)\,f\,{\rm d}f,
\label{eq:qnorm}
\end{equation}
where $A(f)$ is the aperture filter which is an Airy function for circular receiving apertures. The Fresnel filter $S(z,f)$ describes the propagation of the amplitude perturbations of the wave front. In the case of monochromatic light with wavelength $\lambda$,  $S (z,f) = \sin^2(\pi\lambda zf^2)/\lambda^2$. The scale factor of the filter is the Fresnel radius $r_\mathrm{F} = (\lambda z)^{1/2}$\!. The integrand in this formula represents the spatial power spectrum of the scintillation.

Following the Wiener-Khintchine theorem, the covariance of the scintillation between two points separated by a distance $\Delta$ is the result of the Fourier transform of the spatial spectrum. All the previous expressions were averaged over the polar angle $\phi$ since the OT spectrum and all the spectral filters are axi-symmetric. Accordingly, to calculate the $R(z,\theta)$, it is convenient to use the representation of the Fourier transform in polar coordinates: $\tilde X(\Delta) = 2\pi\int X(f)\,f\,J_0(2\pi\Delta f)\,{\rm d}f$, which immediately leads to the expression for the {\it covariance} weighting function:
\begin{equation}
R(z,\theta) = 9.61\int_0^\infty f^{-11/3}S(z,f)\, A(f)\,J_0( 2\pi\theta z f)\,f \,{\rm d}f.\,
\label{eq:rnorm}
\end{equation}
where the distance $\Delta$ was replaced with $z\theta$ from simple geometric considerations.

It can be noted that $R(z,0) = Q(z)$ and the normalised {\it function of the angular correlation} (ACF) in the layer is defined as $\rho(z,\theta) = R(z,\theta)/Q(z)$. For the whole atmosphere, the observed ACF $\varrho(\theta)$ represents a normalised cross-index:
\begin{equation}
\varrho(\theta) = \frac{c(\theta)}{s^2} = \frac{1}{s^2}\int_A C_n^2(z)\,Q(z)\,\rho(z,\theta) \,{\rm d}z.
\label{eq:defth}
\end{equation}

\subsection{Large aperture specifics}
\label{sec:large}

For large telescope with aperture of diameter $D \gg r_\mathrm{F}$, the spectral filter $S(f)$ entering in the integral (\ref{eq:rnorm}) can be simplified:  the filter $A(f)$ overlaps with $S(f)$ a little because the first passes the frequencies below $\sim 1/D$ and the second has main maximum near $1/r_\mathrm{F}$. In the low frequency range, we may replace $\sin^2(\pi\lambda zf^2)$ by $(\pi\lambda zf^2)^2$ without changing the integrand \citep{Roddier1981}. This technique is even more suitable when calculating the covariance because the function $J_0$  suppresses additionally high frequencies.

This simplification leads to important results: 1) the integral dependence on the propagation distance $z$ becomes $\propto z^2$, and 2) the dependence on light wavelength disappears altogether. The latter means that the scintillation becomes almost achromatic, therefore there is no need to use the exact expression of the Fresnel filter \citep{Tokovinin2003} for polychromatic light.

For subsequent calculations, we replace the modulus $f$ by the dimensionless frequency $q = fD$. Then the approximate Fresnel filter assumes the form of $S(z,q) = \pi^2 z^2 D^{-4}q^4$ and the aperture filter changes to $A(q) = [2\,J_1(\pi q)/\pi q]^2$ for clear circular aperture. After these transformations, the integral (\ref{eq:qnorm}) is easily evaluated and is equal to 0.4508, which leads to the usual formula for $Q(z) = 17.34\,D^{-7/3} z^2$.

Most astronomical telescopes have a central obscuration caused by a secondary mirror, for an annular pupil, the aperture filter $A(q)$ is as follows:
\begin{equation}
A(q,\epsilon) = \frac{4}{\pi^2 q^2}\left[\frac{J_1(\pi q)-\epsilon J_1(\pi \epsilon q)}{1-\epsilon^2} \right]^2,
\label{eq:apert_co}
\end{equation}
where $\epsilon$ is equal to the ratio of inner to outer diameters. The use of this aperture filter affects only the numerical factor in the expression for $Q(z)$.

\subsection{Short exposure regime}
\label{sec:assimp}

Here we consider the angular correlation of the scintillation in the case {bf of exposure time} $\tau$ so short that the OT carried by the wind $w$ in the atmosphere does not have time to move across a noticeable distance. For small apertures, the short exposure condition is $\tau \ll r_\mathrm{F}/w$ (the order of few 0.001~s), while for large telescopes, $\tau \ll D/w$  ($\sim 0.01$ or even $\sim 0.1$~s). The latter case is well described by the covariance function (\ref{eq:rnorm}) which becomes, after transition to frequency $q$:
\begin{equation}
R(z,\theta) = 38.4\,D^{-7/3}z^2\!\int_0^\infty \!\!\! q^{-2/3} [J_1(\pi q)]^2 J_0(2\pi \gamma q)\,{\rm d}q,
\label{eq:Rlarge}
\end{equation}
where the integral depends only on the normalised angle $\gamma = \theta z/D$.

\begin{figure}
\centering
\psfig{figure=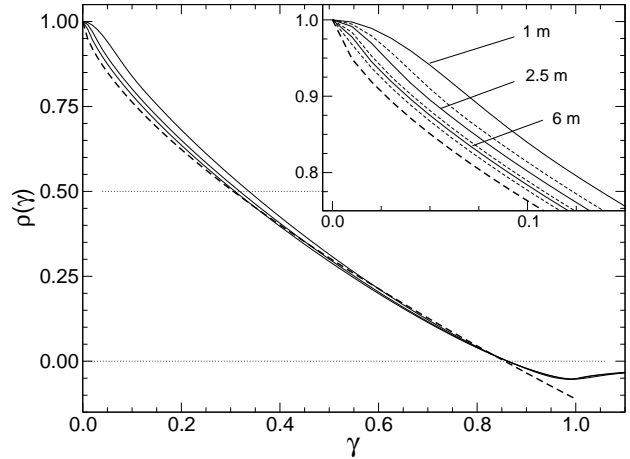,height=8.5cm,angle=-90}
\caption{Angular correlation of the scintillation depending on the normalised angle $\gamma = \theta  z/D$ for short exposure regime. The curves from top to bottom are computed for telescopes diameter of 1~m, 2.5~m and 6~m for turbulence altitude of 15~km (thin lines) and  5~km (thin dashed lines) are shown. The thick dashed line depicts the approximation (\ref{eq:app-large}). In the box, a zoom of the ACFs at low angle is shown. \label{fig:large}}
\end{figure}

For a single layer at distance $z$, the ACF $\rho(z,\theta)$ depends only on $\gamma$ as well:
\begin{equation}
\rho(\gamma) = 2.218\int_0^\infty q^{-2/3} [J_1(\pi q)]^2 J_0( 2\pi \gamma q) \,{\rm d}q,
\label{eq:rho_integral_short}
\end{equation}
and can be calculated analytically through generalised hypergeometric functions (see equation~\ref{eq:rho_short}). A similar result was obtained in \citep{Hickson2004} where the spatial correlation of the scintillation from extended object was considered. There the section of the cone of rays from the Moon played a role of the telescope aperture. In that paper non-normalised functions were calculated, which differ from our expressions by factor 0.4508.

The result of the integration can be approximated by its series expansion:
\begin{equation}
\rho(\gamma) \approx 1 - 1.096\gamma^{2/3} - 0.243\gamma^2 + 0.231\gamma^{8/3}.
\label{eq:app-large}
\end{equation}
This approximation is shown in Fig.~\ref{fig:large}, on which exact functions, calculated by numerical integration of  equation (\ref{eq:rnorm}) with $\lambda = 500\mbox{ nm}$, are also plotted.  One can see that, at $\theta z \lesssim r_\mathrm{F}$, the exact curves are above the approximation due to the simplified Fresnel filter in formula (\ref{eq:Rlarge}).

The value of the ACF at small angle is defined by the diffraction. It is easy to understand this by considering the original expression (\ref{eq:rnorm}) as the convolution of the ACF for an infinitely small aperture with the auto-correlation function of the entrance pupil. The knowledge of the ACF at low angle is essential in problems where a strong correlation is expected. According to Fig.~\ref{fig:large},  the diffraction effect for a 2.5~m diameter telescope can increase the correlation coefficient, e.g., from  $0.9$ to $0.95$ {bf compared to the simplified expression, which corresponds to a two-fold reduction of the variance of the differential signal}.

However, the expansion approximates well the ACF in the domain  $\gamma < 1$. The angle at 50 per cent correlation is obtained from  (\ref{eq:app-large}) for one turbulent layer:
\begin{equation}
\theta_1 = 0.30\frac{D}{z}.
\label{eq:singelarge}
\end{equation}
It is possible to estimate $\theta_1$ by choosing the altitude of the turbulence which produces the main impact on the correlation.

To calculate the ACF $\varrho(\theta)$ related to the whole atmosphere, the formula (\ref{eq:defth}) is to be applied. Assuming $Q(z) \propto z^2$, we obtain that $\varrho(\theta)$ represents the average of the $\rho(z,\theta)$ over all layers weighted by $C_n^2(z)\,z^2$. The angle $\theta_1$, defined as $\varrho(\theta_1) =  1/2$, is, as expected, directly proportional to the telescope diameter.

\begin{figure}
\centering
\psfig{figure=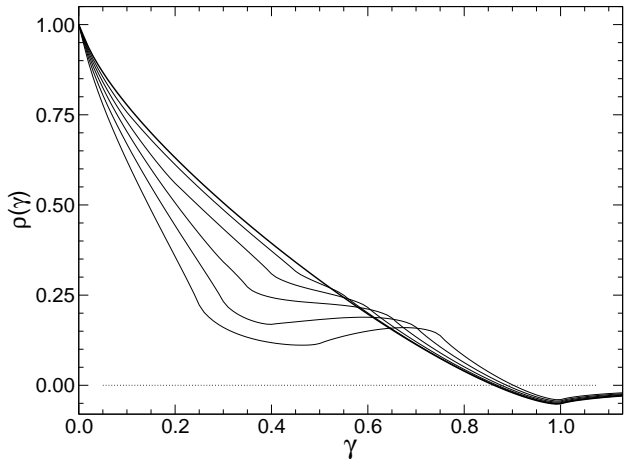,height=8.5cm,angle=-90}
\caption{Short exposure ACFs dependening on the normalised angle $\gamma = \theta  z/D$ for large telescopes with different central obscuration $\epsilon$. The thick line depicts $\epsilon=0$, then $\epsilon$ is equal to 0.1, 0.2, 0.3, 0.4 and 0.5, respectively.
\label{fig:centr-obsc}}
\end{figure}

It is well known that telescope central obscuration affects the scintillation power \citep{Young1967,Dravins1998}. It turns out that it acts upon the angular correlation as well. Fig.~\ref{fig:centr-obsc} shows the  ACFs calculated numerically for different $\epsilon$. For large $\epsilon$, the shape of the curves becomes very complicated. These curves resemble the modulation transfer functions for annular apertures, due to the similarity of their respective mathematical expressions.

The correlation angle for various central obscurations is obtained by multiplying  $\theta_1$ computed for circular aperture by some factor $g(\epsilon)$ listed in Table~\ref{tab:co-coor}. Note that when the inner diameter of the entrance pupil $\epsilon D$ becomes comparable to the $r_\mathrm{F}$ (telescope diameter $\lesssim 0.5$~m), the step on the curves $\rho(\gamma)$ almost disappears.

\begin{table}

\caption{Relative dependence of the angular correlation radius $\theta_1$ on telescope central obscuration $\epsilon$. \label{tab:co-coor}}
\centering
\begin{tabular}{lrrrrrrrrr}
\hline
$\epsilon$      &  0.1  & 0.2  & 0.25 & 0.3 & 0.35 & 0.4 & 0.45 & 0.5 \\
\hline
 $g(\epsilon)$  &  0.95 & 0.81 & 0.72 & 0.67 & 0.62 & 0.57 & 0.52 &  0.47 \\
\hline
\end{tabular}
\end{table}

\subsection{Long exposure regime}
\label{sec:long}

In the previous section the averaging induced by the OT motion during the exposure was negleted. However,  the exposure $\tau$ is usually such that the wind can shift the OT over distances exceeding the telescope aperture size $w\tau \gtrsim D$.

The wind shear of the OT has been introduced in the context of the transition from the spatial correlation to the spatio-temporal correlation \citep{Rocca1974,Caccia1987}. With respect to specific photometric issues, the dependence of the scintillation on the exposure time was studied in \citep{Young1967,Dravins1998,2011AstL}.

In a previous paper \citep{2011AstL} we replaced temporal averaging over the exposure time with spatial averaging over an extended area calculated as the convolution of the aperture $D$ and the wind shear $w\tau$ \citep{Tokovinin2002b}. It is worth noting that in the case of a large wind shear, the ``frozen turbulence'' hypothesis may be not be fully valid due to the temporal evolution of the turbulence. This does not matter in the present case because the spatial spectrum remains unchanged.

The angular correlation of the scintillation in a given direction depends on the relative orientation of the wind $w(z)$. Indeed along the wind direction, the averaged region of the wave front has an area of about $w\tau D$. In the case of transverse displacement at distance $\Delta$, the non-overlapping area $\approx 2\Delta/D$ is much larger than in the case of displacement along the wind $\approx 2\Delta/w\tau$.

Since there is a preferred direction, the problem ceases to be axi-symmetric and a two-dimensional integration over the spatial frequency is required. We carry out this integration in polar coordinates introducing, as dimensionless variables, the frequency $q = Df$, aperture shift $\gamma = \theta z/D$, and wind shear $\omega = w \tau/D$. We rewrite the expression (\ref{eq:rnorm}) in the form of a two-dimensional integral and we add the factor $\sinc^2(\omega q\,\cos\phi)$, which describes the averaging by the wind along the $x$ axis \citep{Tokovinin2002b,2011AstL,2011aMNRAS}:
\begin{multline}
R(z,\theta) = 9.61 D^{5/3}\int_0^\infty q^{-11/3} S(z,q) A(q)\,q\,{\rm d}q\,\\ \times \frac{1}{2\pi}\int_0^{2\pi} \cos(2\pi\gamma q\,\cos(\phi-\psi)) \sinc^2(\pi \omega q\,\cos \phi){\,\rm d}\phi.
\label{eq:wnorm}
\end{multline}
Here $\psi$ is the position angle of the second object in this coordinate system. Note that the function $\sinc \,x = {\sin x}\,/{x}$ is not normalised. Let us denote ${\mathcal L}(q,\omega,\gamma,\psi)$ the integral over the polar angle $\phi$.

The numerical evaluation of the expression (\ref{eq:wnorm}) confirms the strong dependency of the angular correlation to wind direction. However, as the wind can move in any direction relative to the pair of compared stars, the mean effect of wind smoothing is especially important. To evaluate it we may average the expression (\ref{eq:wnorm}) over the angle $\psi$ without loss of significance.

In the integrand of ${\mathcal L}(q,\omega,\gamma,\psi)$, the multiplicand $\sinc$ does not depend on the angle $\psi$, so we can change the order of integration and take it out of the inner integral. The inner integral does not depend on $\phi$ and equals $2\pi J_0(2\pi\gamma q)$. Integration over $\phi$ gives the spectral filter of wind smoothing ${\mathcal T}_1(\omega q)$. An analytic representation is given in \citep{2011AstL}. Finally we have
\begin{equation}
\bar{\mathcal L}(q,\omega,\gamma) ={\mathcal T}_1(\omega q)\,J_0(2\pi\gamma q).
\end{equation}
Since, in the long exposure regime, the ${\mathcal T}_1(\omega q)$ has $1/\pi\omega q$ for asymptote, the average covariance function can be written in the form
\begin{equation}
\bar{\bar R}(z,\theta) = 9.61\frac{D^{8/3}}{\pi w \tau}\int_0^\infty\!\!\! q^{-11/3} S(z,q) A(q) J_0(2\pi \gamma q) {\rm d}q.
\label{eq:aver_wnorm}
\end{equation}
The equation can be simplified further using large aperture specifics (Sect.~\ref{sec:large}) and, for  $\gamma = 0$, turns into expression (7) in \citep{2011AstL}: $Q(z) = 10.66\,D^{-4/3} z^2 (w\tau)^{-1}$. After normalisation, we derive the average ACF $\bar{\bar{\rho}}(z,\gamma)$:
\begin{equation}
\bar{\bar{\rho}}(z,\theta) = 1.15\int_0^\infty q^{-5/3}\left[J_1(\pi q)\right]^2\,J_0(2\pi \gamma q) {\rm d}q.
\label{eq:appver_wnorm}
\end{equation}
The integrand in (\ref{eq:appver_wnorm}) differs from  Sect.~\ref{sec:assimp} in multiplicand $q^{-5/3}$ instead of $q^{-2/3}$. That reduces further the share of high-frequencies in the scintillation spectrum.

We see that the function $\bar{\bar{\rho}}(z,\gamma)$ depends only on the normalised angle $\gamma$  and does not depend on wind shear. Accordingly, the correlation angle $\theta_1$ is also independent on wind speed and exposure time as long as $w\tau \gtrsim D$.

\begin{figure}
\centering
\psfig{figure=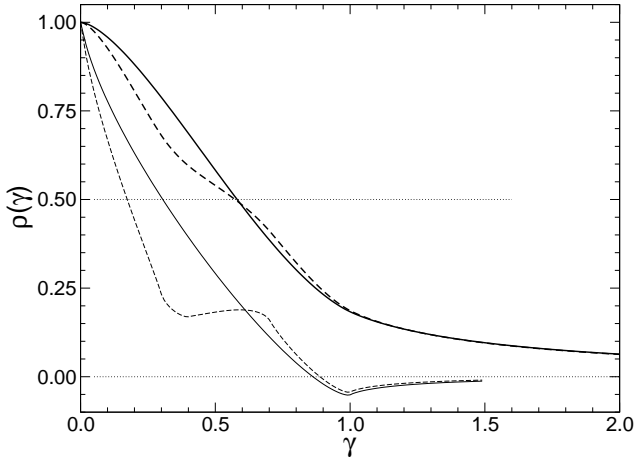,height=8.5cm,angle=-90}
\caption{Long exposure  (thick lines) versus short exposure functions (thin lines) ACFs averaged over all wind directions as a function of the normalised angle $\gamma = \theta  z/D$. The clear circular aperture case is shown with solid lines, the annular aperture case $\epsilon = 0.43$ with dashed ones.
\label{fig:avg-dep}}
\end{figure}

The ACFs $\bar{\bar\rho}(\gamma)$ are shown in Fig.~\ref{fig:avg-dep} in comparison with short exposures functions. The functions become smoother and the effect of the central obscuration becomes less noticeable. It is seen that the angle $\theta_1$ increases by a factor of 1.93 ($\gamma = 0.57$) for clear aperture, and by a factor of 3.35 for a telescope with $\epsilon = 0.43$. It is particularly important that the behaviour near 0 has radically changed. For the correlation of 0.9, the normalised angle is increased from $\gamma = 0.035$ to $\gamma = 0.178$, which means an increase of the angular distance from $1$ to $5$ arcsec for a 2~m class telescope. These numbers are consistent with estimates obtained by \citep{Ryan1998}.

The result of the integration of (\ref{eq:appver_wnorm}) is presented in Appendix~\ref{sec:append},  equation (\ref{eq:rho_long}). A good approximation of the ACF in the range  $\gamma < 1$ can be obtained using its series expansion:
\begin{equation}
\rho(\gamma) \approx 1 - 2.79\,\gamma^{5/3} + 1.78\,\gamma^2 + 0.311\,\gamma^{11/3} -0.141\,\gamma^4.
\end{equation}

The previous equations are valid for clear circular apertures, but Fig.~\ref{fig:avg-dep} shows that the introduction of a central obstruction does not change the value of the angle giving 1/2 in the long exposure regime.

\section{Prediction of the angular correlation of the scintillation}
\label{sec:prognoz}

We apply the above development to the vertical profiles of the OT monitored with the MASS/DIMM instrument \citep{2003SPIE,2007bMNRAS} at Mt Shatdzhatmaz (2127~m above sea level) in the North Caucasus \citep{2010MNRAS} to estimate the angular correlation for the whole turbulent atmosphere.

The database for the period November 2007 -- October 2009 consists of $\approx 85\,000$ profiles. Each profile was obtained as a result of one-minute measurement and the restoration was performed on a 13-layers grid composed of the surface layer and 12 layers centered at 0.5, 0.71, 1.0, 1.41, 2.0, 2.82, 4.0, 5.66, 8.0, 11.3, 16.0 and 22.6~km. The output of the MASS profile restoration is, for each altitude, the intensity $\Delta J_i$ equal to the integral of the index of refraction structure coefficient within this layer: $C_n^2(z_i)\delta z$.

During restoration, the top node of the grid (22.6~km) accumulates the noise of the inversion process, and its output intensity may be somewhat overstated. For this reason, we decrease by half the intensity of this layer in further estimates. The effect of such a correction does not exceed 10 per cent.

The results of the simulations should be used in planning of future high-precision photometric observations and in the design of the SAI 2.5~m telescope equipment. That is exactly why in this paper we often use parameters of this particular telescope: entrance aperture diameter $2.5$~m and central obscuration $\epsilon = 0.43$.

\subsection{Angular correlation in the short exposure regime for the SAI 2.5~m telescope}
\label{sec:short_exposure}

\begin{figure}
\centering
\psfig{figure=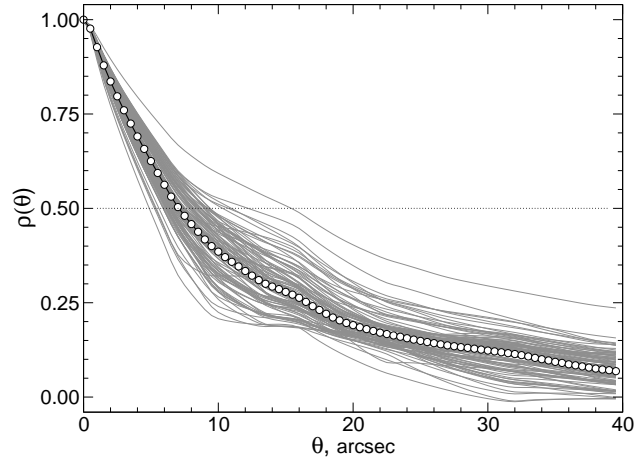,height=8.5cm,angle=-90}
\caption{Example of the short exposure ACFs computed for real OT vertical profiles for the SAI 2.5~m telescope. 100 randomly chosen records are drawn, the bright dots denote the median  of 1684 individual records. \label{fig:samples-large}}
\end{figure}

In order to predict the angular correlation of the scintillation on the SAI 2.5~m telescope with short exposures, the corresponding covariance function was calculated for each OT profile by the formula
\begin{equation}
c(\theta) = \sum_{i=0}^{12} z_i^{2}\Delta J_i \rho(\theta z_i/D),
\label{eq:comp2}
\end{equation}
and then normalised by $c(0)$ to obtain the ACF $\varrho(\theta)$ and $\theta_1$, the angle at which $\varrho(\theta_1) = 1/2$. We used the exact ACF in the layers $\rho(\theta z_i/D)$, calculated for each node $z_i$ by the formula (\ref{eq:rnorm}), to take into account the central obscuration and diffraction effects. Fig.~\ref{fig:samples-large} shows randomly selected resulting ACFs. The median ACF for enlarged sample is plotted as well, showing the characteristic step due to central obscuration. Also note the long ($30 - 40$ arcsec) pedestal attributed to the presence of low-altitude turbulence.

\begin{figure}
\centering
\psfig{figure=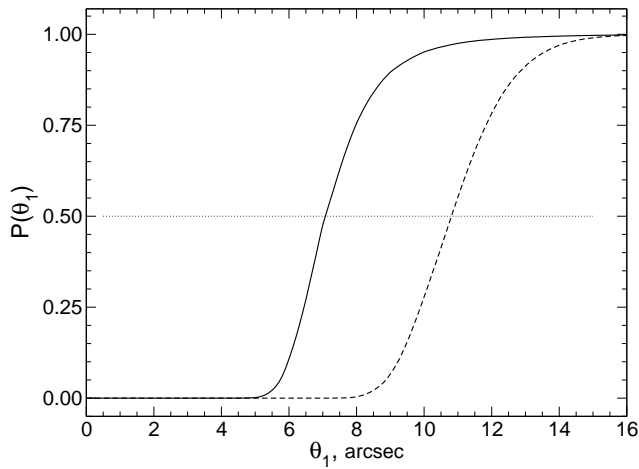,height=8.6cm,angle=-90}
\caption{Cumulative distributions of the correlation angle $\theta_1$, predicted for short exposures on the SAI 2.5~m telescope. The solid and dashed lines correspond respectively to the distributions for the telescope and for a clear circular 2.5~m aperture.\label{fig:iso-large}}
\end{figure}

The median of the $\theta_1$ distribution on Fig.~\ref{fig:iso-large} is equal to $7.1$ arcsec. The first and third quartiles of the distribution are equal to $6.4$ and $8.0$ arcsec, respectively. Replacing $\theta_1/D$ ratio in formula (\ref{eq:singelarge}), we estimate the typical effective altitude of the OT, after correction for the central obscuration, to $\approx 12$~km.

\subsection{Angular correlation in the long exposure regime for the SAI 2.5 m telescope}
\label{sec:long_exposure}

Although, the normalised averaged ACF does not depend on wind, the weight of each turbulent layer does because the scintillation power is proportional to $(w \tau)^{-1}$ for long exposures. Therefore, in addition to the OT profiles, some information is required on the vertical distribution of wind speed. For the prediction, we used the vertical wind profile $\{V(z_i)\}$ provided by the NCEP/NCAR meteorological model \citep{NCEP-NCAR} averaged during the observation period.

The ``observed'' covariance function was constructed using the formula
\begin{equation}
c(\theta) = \sum_{i=0}^{12} z_i^{2}\frac{\Delta J_i}{V_i} \bar{\bar\rho}(\theta z_i/D),
\label{eq:comp3}
\end{equation}
where  $\bar{\bar\rho}(\theta z_i/D)$ in each layer was calculated using the exact formula (\ref{eq:aver_wnorm}). Then, as in the previous case, the function was normalised and the angle $\theta_1$ was determined.

\begin{figure}
\centering
\psfig{figure=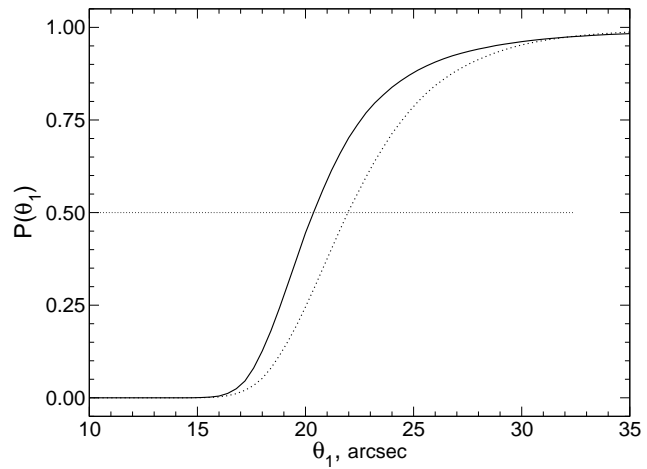,height=8.6cm,angle=-90}
\caption{Cumulative distributions of the correlation angle $\theta_1$, predicted for long exposures on the SAI 2.5~m telescope. The dashed line shows the case without considering wind profile.
\label{fig:long-distr}}
\end{figure}

The cumulative distribution of $\theta_1$ for all OT profiles is shown in Fig.~\ref{fig:long-distr}. Compared to the short exposures case, the correlation angle is significantly larger with a median of $20.3$ arcsec, and the first and third quartiles of $19$ and $23$ arcsec, respectively. The distribution is very narrow, and these numbers can be used with confidence for planning the observations. Neglecting the wind distribution changes the correlation angle by $\approx 10$ per cent only.

\begin{figure}
\centering
\psfig{figure=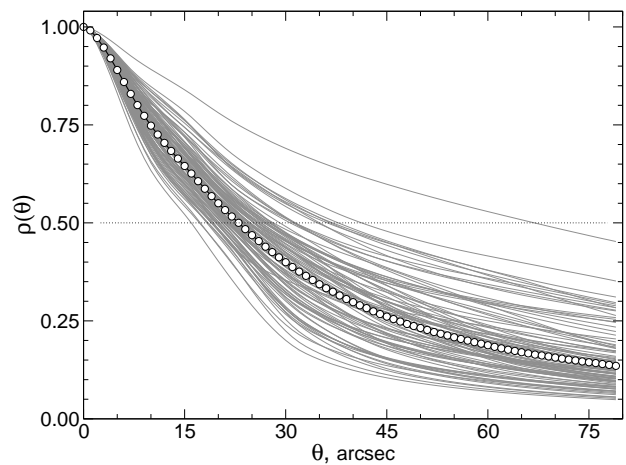,height=8.5cm,angle=-90}
\caption{Example of the long exposure ACFs computed by real OT vertical profiles for the SAI 2.5~m telescope. 100 randomly chosen records are shown, the bright dots denote the median of 1618 individual records.
 \label{fig:long-samples}}
\end{figure}

Examples of individual ACFs and their median function are shown in Fig.~\ref{fig:long-samples}. As in the previous section, the median ACF is computed as median values of individual functions for fixed $\theta$ and respective angle $\theta_1$ is computed for $\varrho(\theta) = 1/2$. This procedure leads to a slightly different result from the simple median of all the individual $\theta_1$.

In general, the median ACF is similar to the function for a single layer (Fig.~\ref{fig:avg-dep}). Features near $\theta = 0$ are defined by the high and quite stable turbulence. A strong correlation of 0.9 is observed for objects at an angular distance of $\approx 5$ arcsec. When $\theta \approx 10$ arcsec, the correlation coefficient can be about 0.75, which corresponds to twofold reduction of the scintillation noise with respect to the distance equal $\theta_1$ and fourfold compared to the case of far apart objects.

For this photometric regime, the effective altitude $h_\mathrm{eff} = 0.57\,D/\theta_1$ is $14\mbox{ km}$, which corresponds to a barometric altitude of 16~km.

\section{Analysis of the scintillation noise reduction methods}
\label{sec:scnoise-dec}

\subsection{Method of conjugated photometry}
\label{sec:conj}

\citet{Osborn2010} proposed a method for differential photometry where the entrance pupil of the telescope is conjugated to the most intense turbulent layer. According to the authors, this operation increases the correlation of the scintillation and consequently reduces the scintillation noise in the differential signal. Such a geometry of light propagation increases the contribution of the low-altitude turbulence to the scintillation but the gain in the correlation is expected to be larger.

However real OT vertical distributions may be substantially different from the simplified model used in the paper. First of all, the maximum contribution in the scintillation is not always produced by the most intense layer. Because of the $z^2$ weighting factor, higher layers with weaker OT can dominate. Secondly, in the case of short exposures,  it was shown that the correlation is reduced very quickly even for small angular separation. This means that even a weak high-altitude turbulence can limit strongly correlated scintillation to a very narrow region of the sky. For the case of long exposures the ACF peak is broader but a role of slowly moving OT in the surface layer grows up. In any case, an amplification of the scintillation, caused by the transfer of the entrance pupil to the conjugate altitude $z^*$, is very significant.

To analyse the situation, we define $\theta_c$ as the angle at which {bf conjugated photometry decreases the scintillation noise power by a factor of 2} compared to conventional differential photometry (assuming uncorrelated scintillation):
\begin{equation}
s^2 = 2\sigma^{*2} (1-\varrho^*(\theta_c)) \qquad \mbox{or} \qquad \varrho^*(\theta_c) = 1 - \frac{s^2}{2\sigma^{*2}},
\end{equation}
where $s^2$ is the scintillation index under a given state of the atmosphere, $\sigma^{*2}$ and $\varrho^*(\theta)$ are the scintillation power and the ACF for the whole atmosphere in the case of the conjugated pupil. Using the expression for the scintillation in a case of virtual propagation \citep{Fuchs1998} and our data on the OT vertical distribution, we calculated the correlation as
\begin{equation}
\varrho^*(\theta) = \frac{1}{\sigma^{*2}}\sum_{i=0}^{12} \Delta J_i (z_i-z^*)^{2} \rho(\theta z_i/D).
\label{eq:comp_conj}
\end{equation}
Evidently, for objects closer than $\theta_c$, the gain in noise suppression is greater than 2. Computing $\rho(\theta z_i/D)$ we did not take into account the loss of a part of the telescope aperture in  conjugated pupil mode. Increasing the central obscuration additionally amplifies the scintillation and reduces the correlation.

\begin{figure}
\centering
\psfig{figure=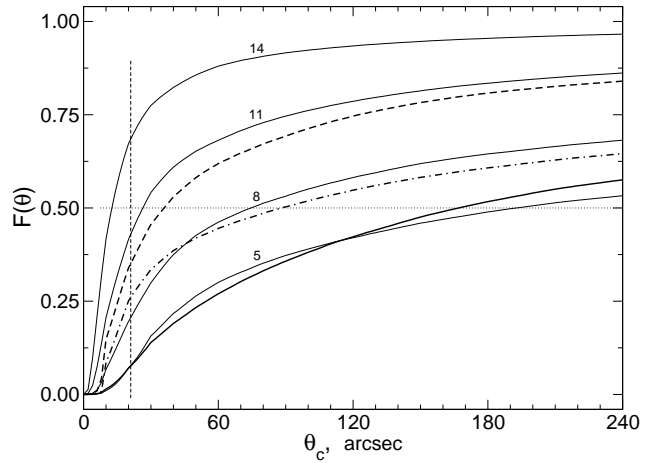,height=8.5cm,angle=-90}
\caption{Distribution of the angles $\theta_c$ calculated with real OT profiles for different conjugation altitudes $z^*$ (thin lines) in km. The thick solid line depicts the adaptive selection of the $z^*$ minimising the scintillation in the free atmosphere, the dash-dotted line corresponds to the adaptive selection of the most intensive layer. The thick dashed line is the same as solid, but for the scintillation noise reduction by 3 times. The vertical line corresponds to $\theta_1$.
\label{fig:conj-short}}
\end{figure}

Distributions of these estimates are presented in Fig~\ref{fig:conj-short} for short exposures. Fixed $z^*$ distributions show that in our conditions (when the surface layer contains $\approx0.65$ of total turbulence), the conjugation to 5~km altitude is more profitable because it minimises the scintillation from the surface turbulence. Thus, when $z^* = 5\mbox{ km}$, the median  $\theta_c$ reaches $190$ arcsec, and in a quarter of the time  $\theta_c > 20$ arcmin. A similar picture is observed in the case of an adaptive choice of $z^*$, minimising the scintillation in the free atmosphere ($z \ge 1\mbox{ km}$). Results are worse when conjugating to the most intense free atmosphere layer.

The distributions show the ``toggle'' character of the process: in the case of strong surface turbulence, the angles $\theta_c$ are small and the application of the method is meaningless. But above a certain level, the technique becomes effective. An attempt to reduce the scintillation noise more (by 3 times) is shown by the dashed line: an adaptive choice $z^*$ provides the median angle $\approx 35$ arcsec only.

In the long exposures case the main characteristics of the distributions are preserved but the correlation angles become smaller. E.g., for the adaptive choice of the conjugation altitude, the median $\theta_c$ is $\approx 40$ arcsec, and $\theta_c > 2$ arcmin occurs only in 18 per cent of the time. The main reason for the deterioration in conjugated mode is an increase of the scintillation due to slow low-altitude turbulent layers.

For a successful application of the discussed technique, data on the current OT and wind above the telescope are essential since these characteristics are very unstable. A study of the temporal evolution of  $\theta_c$ shows that it can change by several times over the time scale of tens of minutes.  Note, that the altitude $z^*$, providing the minimum amplification of the scintillation, coincides with the effective altitude of the turbulence in free atmosphere. The software of the MASS/DIMM instrument \citep{2003MNRAS} delivers this parameter in real time.

\subsection{Ensembles}
\label{sec:ensembles}

Differential CCD photometry using a close group of stars, located around the object of interest, was proposed primarily to minimise the impact of variations of the atmospheric transparency in order to obtain high-precision (better 0.001 magnitude) data for the study of stellar micro-variability \citep{Brown1994,Heasley1996}. This technique was also found useful for observation of systems with exoplanets \citep{Everett2001,Southworth2009,Mann2011}.

The {\it ensemble of the comparison stars} should be fairly compact as the spatial correlation of the fluctuations in the extinction caused by, e.g., cirrus, has a radius no more than a few arcminutes. This issue is poorly understood, but every astronomer can easily verify this by looking at clouds patterns passing over the Moon.

It was also noted that the ensemble reduces the contribution of the stellar scintillation to the error budget of the measurements \citep[see, e.g.,][]{Gilliland1993}. We investigate in this section the potential of this technique invoking the angular correlation of the scintillation. First we discuss the procedure for comparing the measured signal.

There are two obvious ways to compare the flux from the object $I_*$ to the total flux $I_\Sigma$ of $N$ stars in the ensemble:
\begin{equation}
o = \frac{I_*}{I_\Sigma}, \mbox{  where  } I_\Sigma = \sum\limits_j^N I_j,
\label{eq:intens}
\end{equation}
and/or alternatively, in the space of magnitudes
\begin{equation}
\Delta m = m_* - \frac{1}{N}\sum_j^N m_j
\label{eq:mags}
\end{equation}
In general, to each star in the ensemble can be attributed a certain weight. Evidently, the formula (\ref{eq:intens}) shows the contribution of multiplicative noise for fainter stars with less weight.

We write the analogue of the formula (\ref{eq:mag_index}) for the ensemble:
\begin{equation}
\frac{\sigma^2_{o}}{o^2} = s^2\Bigl[1 - 2\!\sum_j w_j\varrho(\theta_{*,j}) + \sum_j w_j^2 + 2\!\sum_{k,j<k} w_jw_k\varrho(\theta_{j,k})\Bigr],
\label{eq:intens2}
\end{equation}
where $w_j = I_j/I_\Sigma$ is the relative weight of $j$-th star in the fluxes. In the magnitudes space, all the stars in the ensemble are equally accurate, and
\begin{equation}
\frac{\sigma^2_{\Delta m}}{1.179} = s^2\Bigl[1\!-\!\frac{2}{N}\sum_j \varrho(\theta_{*,j})\!+\!\frac{1}{N}\!+\! \frac{2}{N^2}\sum_{k,j<k}\varrho(\theta_{j,k})\Bigr].
\label{eq:mag2}
\end{equation}
In these expressions, the sum in brackets can be interpreted as {\it an amplification} of the scintillation noise. The second term in the sum describes the correlation of the star under study with the comparison stars, the third and fourth term are the variance of the mean brightness of correlated comparison stars.

For clarity, we are going to analyse the latter formula. If the distances between all the stars are much larger than $\theta_1$, the resulting variance is reduced to $s^2(1 + N^{-1})$, i.e. an increase of the number of the stars reduces the scintillation noise. However, if close pairs appear in the ensemble, further improvement is slowed down by the fourth term. Nonetheless in the limit, when the ensemble covers the sky area much larger than $\theta_1^2$, it can be regarded as an extended object, which has no scintillation.

When the ensemble of stars are at distances from the object comparable to the radius of the angular correlation, the second term, the average correlation coefficient, is to be taken into account. Winning situation is achieved when the object under study is located in the centre of ensemble and is surrounded by a large number of comparison stars.

As a numerical example, let us consider the ensemble of $N$ stars arranged in a circle of radius $R$ around the star under study. The amplification for different configurations is given in Table~\ref{tab:constell}. The median curve from Section~\ref{sec:long_exposure} is used as the ACF with the angle $\theta_1 = 23$ arcsec.

\begin{table}
\caption{Amplification of the scintillation noise for different configurations of star ensemble. See text for explanation. \label{tab:constell}}
\centering
\begin{tabular}{lrrrrrrr}
\hline
$R$         & $N=1$ & 2 & 3 & 4 & 6 & 8 &  $\infty$ \\
\hline
$0.5\,\theta_1$    & 0.57 & 0.32 & 0.27 & 0.26 & 0.24 & 0.24 &  0.24 \\
$1\,\theta_1$      & 1.00 & 0.63 & 0.53 & 0.50 & 0.48 & 0.47 &  0.46 \\
$2\,\theta_1$      & 1.49 & 1.05 & 0.91 & 0.85 & 0.80 & 0.79 &  0.77 \\
$4\,\theta_1$      & 1.78 & 1.30 & 1.15 & 1.08 & 1.01 & 0.98 &  0.93 \\
\hline
\end{tabular}
\end{table}

The first row of the table reflects the situation of a crowded stellar field when the typical distance between the stars ($12$ arcsec in our example) is less than the angle of the correlation. We see that in this case there is no reason to have more than 4 comparison stars. Further increase in their numbers do not improve the situation even if they are close to the star. Adding new stars at larger distance, we reduce the second term in (\ref{eq:mag2}) and increase the amplification.

The last line describes the case of rarefied stellar ensemble, the typical distance between stars is $1.5$ arcmin. When using a single comparison star, the amplification factor is $\approx 1.8$. However, we can reduce its value up to 1, corresponding to a differential measurement at distance $\theta_1$, by choosing 6 to 8 comparison stars.

Calculations of the scintillation power from OT profiles, give the median scintillation noise equal to $250~\mu$mag for 60~s integration on the SAI 2.5~m telescope. It means that the scintillation noise can be as small as $120~\mu$mag using the ensemble technique, to be compared to $350~\mu$mag with trivial differential photometry.

\section{Discussion}
\label{sec:discuss}

All the equations in the paper are written for the distance $z$ to turbulent layer and not for its altitude $h$ above the observatory. The dependency of the scintillation power on the zenith distance $Z$ (or air mass $M_\mathrm{Z} = \sec Z$) is obtained by replacing $z$ at $h\sec Z$ in the formulae (\ref{eq:s2int}) and (\ref{eq:cint}) what leads to $\propto M_\mathrm{Z}^{3}$. In the long exposure regime, the exponent varies from 3 to 4, depending on the wind direction \citep {Young1967,2011AstL}. 

In any case, this is a common factor for all the layers and it does not change their relative contribution to the scintillation. Therefore, to identify the dependency of  $\theta_1$ on air mass, we can use the formula (\ref{eq:singelarge}) which implies $\theta_1 \propto M_\mathrm{Z}^{-1}$. Thus, for differential photometry far from the zenith, not only the scintillation power greatly increases, but also the correlation of the scintillation decreases. In the two limiting cases, when the sources are either very close ($\varrho \approx 1$) or very far ($\varrho \approx 0$), the variance of the differential signal increases together with the $s^2$. However, for distances of order $\theta_1$ ($\varrho \approx 0.5$),  $\sigma^2_{\Delta m}$ grows faster.

We did not touch the absolute intensity of the scintillation, analysing the comparative effects, all other conditions being equal. The atmosphere above the tropopause is generally stable enough, however a growth of $s^2$ occurs in presence of strong turbulent layers at the tropopause altitude (jet streams) or below in form of wind wake from mountain ridges or severe weather phenomena. In such cases, the angle of the correlation should also increase. Similar dependence is observed in our data, but it is very blurred and $\theta_1$ changes only at $\sim 50$ per cent when the $s^2$ varies by a few times.

Since the typical correlation angle corresponds to a few meters in the high atmosphere, a deviation of the spatial spectrum of the refractive index from the Kolmogorov model is to be taken into account. It is known that the finite outer scale of the non-Kolmogorov turbulence greatly reduces the spatial spectrum already in a meter range, what affects a size of image at large telescopes \citep{Martinez2010}.

In our case, this factor suppresses low-frequencies scintillation what is to lead to a shortening of the correlation distance. This effect resembles the central obscuration effect which amplifies the high-frequencies. The outer scale at high altitudes is large, with a characteristic length about 20 to 50~m \citep{DaliAli2010}. This gives reason to assert that the obtained results are applicable for telescopes 2 -- 4~m class with confidence.

The comparison of photometric methods to reduce the scintillation noise: ensemble of comparison stars and conjugate-plane photometry, leads to the following conclusions. The first method requires a close group of stars which can not always be found. But if there is a compact group of 4 -- 5 stars, a reduction of the scintillation noise is possible up to 3 -- 4 times. The method is always applicable because is associated to the high-altitude turbulence. With larger telescopes, more extended sky area can be used.

The feasibility of conjugate photometry depends on the turbulence in the lower atmosphere. Usually it is distinguished by high instability. Our simulations show that in the sites where low-altitude OT normally dominates, threefold gain in signal-to-noise ratio is not always attainable. But in certain atmospheric conditions, the comparison star can be chosen far from the target star ($2 - 4$ arcmin). The question is how to predict and detect these conditions?

\section{Conclusion}
\label{sec:conclusion}

In this paper, we discussed the problem of correlation of the stellar scintillation from two point-like sources, closely spaced in the sky. Using the standard model of the optical turbulence (Kolmogorov spatial spectrum of refractive index fluctuations) in the  approximation of the weak perturbations, two situations were considered for large telescope aperture (much larger than the Fresnel raduis $r_\mathrm{F}$): 1) observations with short exposures, 2) observations with long exposures (typical photometric practice).

Analytic expressions obtained for the ACFs, have lengthy description, therefore a simple power-law approximation, well describing the functions, has been proposed. As the {\it measure} of the angular correlation, the angle  $\theta_1$, at which the correlation coefficient is equal 1/2, was adopted.

Based on previously obtained data on the vertical distribution of the OT in the atmosphere above Mt Shatdzhatmaz in the Northern Caucasus, the simulation experiments have been performed for the SAI 2.5~m telescope in order to predict the angular correlation of the scintillation for photometric observations at this observatory. All hereinafter referred numerical results relate to this telescope and this site.

Since in case of a large telescope, the ACF depends only on one parameter: the normalised angle $\gamma = \theta\,z/D$, it is possible to scale the obtained estimates for telescope with different diameter. Central obscuration in a telescope affects the correlation of the scintillation for short exposures but not for long exposures.

Simulation for short exposure regime (fast photometry) predicts the median of the $\theta_1$  as large as $7$ arcsec. In the long exposure regime, the ACF averaged over all wind directions, does not depend on  wind speed and exposure. In this case, estimated median of the correlation angles $\theta_1$ amounts to $\approx 20$ arcsec.

Study of the conjugate-plane photometry shows that the method depends critically on the intensity of low-altitude OT. When the lower atmosphere is calm, one can use the comparison star at distance greater than several arcminutes. Analysis of the ensemble photometry shows that the compact, but quite resolved, ensemble of four comparison stars is enough to get the {reduction in the scintillation noise power by} $\approx 8$ times. Photometric observation of the different ensemble configurations could provide the necessary experimental verification of the proposed formulae and the performed simulations.

\section{Acknowledgements}

The author is grateful to his colleagues, actively participated in obtaining data on optical turbulence at Mt Shatdzhatmaz and personally for A.~Tokovinin and B.~Safonov for useful discussion and M.~Sarazin for numerous comments. In a sense, the work is a continuation of the paper \citep{2011aMNRAS} and the author thanks the unknown referee, who  has further stimulated this research by his remark to the previous paper.

\bibliography{isoplan_mn}
\bibliographystyle{mn2e}

\appendix

\section{Exact formulae for ACFs}
\label{sec:append}

\subsection{Short exposures}

In the case $D \gg r_\mathrm{F}$ (irrelevance of diffraction effect) without central obscuration in the short exposure regime, the integration of the (\ref{eq:rho_integral_short}) gives:
\begin{align}
\rho(\gamma) &= {}_3F_2\left({\textstyle -\frac{5}{6},\frac{1}{6},\frac{7}{6}};{\textstyle \frac{2}{3},1};\gamma^{2} \right) - \notag\\ &- D_1\gamma^{2/3}\,{}_3F_2\left({\textstyle -\frac{1}{2},\frac{1}{2},\frac{3}{2}};{\textstyle\frac{4}{3},\frac{4}{3}};{\gamma}^{2}\right), \quad &(\gamma < 1) \notag\\
\rho(\gamma) &= D_2\gamma^{-7/3}\,{}_3F_2\left({\textstyle\frac{7}{6},\frac{7}{6},\frac{3}{2};2,3;\gamma^{-2} }\right), \quad &(\gamma > 1)
\label{eq:rho_short}
\end{align}
where
\begin{align}
D_1 &= \frac{45\sqrt{3}\,2^{2/3}}{8\pi^3}\Gamma\left(\textstyle\frac{2}{3}\right)\Gamma^4\left(\textstyle\frac{5}{6}\right) = 1.0965, \notag \\
D_2 &= -\frac{5\sqrt{3}}{288}\,\frac {\Gamma{\textstyle\left(\frac{5}{6}\right)} \Gamma{\textstyle\left(\frac{2}{3}\right)}}{\sqrt {\pi }} = -0.0259.\notag
\end{align}

\subsection{Long exposures}

In the case of large telescope without central obscuration in the long exposure regime, the integration of the (\ref{eq:appver_wnorm}) results:
\begin{align}
\rho(\gamma) &=
{}_3F_2\left({\textstyle -\frac{4}{3},-\frac{1}{3},\frac{2}{3}};{\textstyle\frac{1}{6},1};{\gamma}^{2} \right) - \notag \\
&- E_1\,\gamma^{5/3}\,{}_3F_2 \left({\textstyle-\frac{1}{2},\frac{1}{2},\frac{3}{2}};{\textstyle\frac{11}{6},\frac{11}{6}};\gamma^{2}\right), \quad &(\gamma < 1) \notag\\
\rho(\gamma) &= E_2\,\gamma^{-4/3}\,{}_3F_2\left({\textstyle\frac{2}{3},\frac{2}{3},\frac{3}{2}};2,3;{\gamma}^{-2} \right), \quad &(\gamma > 1)
\label{eq:rho_long}
\end{align}
where
\begin{align}
E_1 &= {\frac {64}{225}}\frac{ {\pi }^{2}\sqrt [3]{2}\sqrt{3}}{\Gamma^4\left(\frac{5}{6}\right)\,\Gamma\left(\frac{2}{3}\right)} = 2.7867, \notag \\
E_2 &= {\frac {2}{81}}\,\frac{ \pi^{3/2}\sqrt{3}}{\Gamma
\left(\frac{5}{6}\right)\Gamma\left(\frac{2}{3}\right)} = 0.1558.\notag
\end{align}

\label{lastpage}

\end{document}